\documentclass[
    ,final            
  ]
  {aipproc}

\layoutstyle{8x11single}

\usepackage{bm}        
\usepackage{amssymb}   
\usepackage{amsmath}   
\usepackage{tabularx}   

\begin{document}

\title{The Bayesian reconstruction of the in-medium heavy quark potential from lattice QCD and its stability}


\author{Yannis Burnier}{
  address={Institut de Th\'eorie des Ph\'enom\`enes Physiques, Ecole Polytechnique F\'ed\'erale de Lausanne, CH-1015, Lausanne, Switzerland}
}

\author{Olaf Kaczmarek}{
  address={Fakult\"at f\"ur Physik, Universit\"at Bielefeld, D-33615 Bielefeld, Germany}
}

\author{Alexander Rothkopf}{
  address={Institute for Theoretical Physics,  Heidelberg  University, Philosophenweg 16, D-69120 Heidelberg, Germany}
}

\begin{abstract}
We report recent results of a non-perturbative determination of the static heavy-quark potential in quenched and dynamical lattice QCD at finite temperature. The real and imaginary part of this complex quantity are extracted from the spectral function of Wilson line correlators in Coulomb gauge. To obtain spectral information from Euclidean time numerical data, our study relies on a novel Bayesian prescription that differs from the Maximum Entropy Method. We perform simulations on quenched $32^3\times N_\tau$ $(\beta=7.0,\xi=3.5)$ lattices
with $N_\tau=24,\ldots,96$, which cover $839{\rm MeV} \geq T\geq 210 {\rm MeV}$. To investigate the potential in a quark-gluon plasma with light u,d and s quarks we utilize $N_f=2+1$ ASQTAD lattices with $m_l=m_s/20$ by the HotQCD collaboration, giving access to temperatures between $286 {\rm MeV} \geq T\geq 148{\rm MeV}$. The real part of the potential exhibits a clean transition from a linear, confining behavior in the hadronic phase to a Debye screened form above deconfinement. Interestingly its values lie close to the color singlet free energies in Coulomb gauge at all temperatures. We estimate the imaginary part on quenched lattices and find that it is of the same order of magnitude as in hard-thermal loop perturbation theory. From among all the systematic checks carried out in our study, we discuss explicitly the dependence of the result on the default model and the number of datapoints. 
\end{abstract}

\maketitle


\section{Introduction}

Nuclear matter under extreme conditions provides ample challenges to our understanding of the mechanism of confinement based on microscopic quantum-chromo dynamics. Strong fluctuations, be it at the high temperatures that existed shortly after the big bang or at the high densities in the interior of neutron stars, lead to an emergence of partonic degrees of freedom, confined in hadrons at everyday energy scales. With the advent of modern collider facilities, such as RHIC and LHC, as well as future facilities, such as FAIR, it is possible to recreate these scenarios directly in the laboratory.

To gain quantitative insight into the dynamical aspects of deconfinement, the bound states of a heavy quark and antiquark are ideal probes \cite{Brambilla:2010cs}. The ground states of these two-body systems in vacuum possess lifetimes of the order of inverse keV \cite{Agashe:2014kda}, which makes them essentially stable from the point of view of QCD. If immersed in a thermal bath they are thus able to accumulate effects from the medium over time, while still remaining well defined particles that can be cleanly measured in experiment. Eventually, as temperature is increased, they will melt into their constituents. How and over which timescales this melting occurs, has been and remains the focus of a wealth of theoretical studies (for a selection of lattice QCD based studies to in-medium modification of heavy quarkonium see \cite{Umeda:2002vr,Asakawa:2003re}).

A first principles understanding of the melting process in turn opens up the possibility to extract from the measured yields in relativistic heavy-ion collisions \cite{Adare:2006ns} the properties of the medium created in the collision center. This essentially would turn heavy quarkonium into a dynamical thermometer for RHIC and LHC. While the classic idea of a suppression of yields in the presence of a QGP \cite{Matsui:1986dk} invigorated this area of research, we now know that an intricate interplay between melting and regeneration \cite{Liu:2009nb}, needs to be understood before quantitative predictions can be made.

What distinguishes heavy quarkonium, such as e.g. bottomonium (the $b\bar{b}$ vector channel), from lighter mesons is the inherent separation of scales between the constituent quark mass and the medium temperature, as well as the characteristic scale of QCD ($4.66{\rm GeV}=m_b\gg \Lambda_{\rm QCD}\simeq 200{\rm MeV}$). From the rough estimates for the initial temperature based on direct photons at RHIC ($T_{\sqrt{s}=200{\rm MeV}}\simeq221$MeV) \cite{Adare:2008ab} and LHC ($T_{\sqrt{s}=2.76{\rm GeV}}\simeq305$MeV) \cite{Wilde:2012wc} we see that indeed $m_b\gg T$. Intuitively this tells us that under the conditions of current heavy-ion experiments, neither quantum nor thermal fluctuations can spontaneously create a heavy quark pair, so that a description in terms of non-relativistic physics should be feasible. Note that these temperatures are still close to the deconfinement temperature $T_c\simeq 160$MeV as determined from lattice QCD \cite{Aoki:2006we,Bazavov:2011nk} and hence the medium partons remain strongly interacting.

Using effective field theories (EFT) \cite{Brambilla:2004jw} it has been shown that the effects of the interactions between a static quark-antiquark pair and a thermal medium can be summarized in a Schr\"odinger equation with an effective in-medium potential. The derivation of the potential relies on a matching procedure to fundamental QCD, and relates its values to the late real-time behavior of the rectangular Wilson loop
\begin{align}
V(r)=\lim_{t\to\infty} \frac{i\partial_t W(t,r)}{W(t,r)}\label{Eq:VRealTimeDef}, \quad W(t,r) = \left\langle {\rm exp}\Big[ - {ig}\int_{\square} dx_\mu A^\mu(x) \Big] \right\rangle .
\end{align}
A first evaluation of this quantity in hard-thermal loop perturbation theory \cite{Laine:2007qy} showed that the potential at finite temperature is complex valued. As expected at temperatures far above $T_c$, the presence of deconfined partons leads to Debye screening of the real part. The imaginary part on the other hand has been attributed to Landau damping (scattering with medium gluons) \cite{Beraudo:2007ky} and a subsequent systematic EFT analysis revealed \cite{Brambilla:2008cx} additional contributions from the transition of a singlet to an octet state (absorption of medium gluons)\footnote{The presence of the imaginary part has been given a dynamical interpretation in the language of open-quantum systems \cite{Akamatsu:2011se}, where it is related to the remnants of thermal medium fluctuations, which were integrated out when establishing the effective potential description}.

Before the advent of a modern effective field theory for in-medium heavy quarkonium, the temperature dependence of the real-part of the static inter-quark potential has often been modeled in phenomenological studies. Some studies used the color singlet free energies $F^{(1)}(r)$ others the internal energies $U^{(1)}(r)$ or even linear combinations thereof \cite{Nadkarni:1986as}. Interestingly, resummed HTL perturbation theory to leading order tells us that ${\rm Re}[V]$ and $F^{(1)}$ agree but this relation already appears to break down at next to leading order \cite{Burnier:2009bk}. To provide phenomenology with dependable input, the question of how to evaluate Eq.\eqref{Eq:VRealTimeDef} in a non-perturbative manner becomes a pressing issue.

Simulations of lattice QCD provide us with non-perturbative access to the static properties of QCD at finite temperature, e.g. the equation of state \cite{Borsanyi:2013bia}. These calculations are carried out in Euclidean time, so that real-time quantities, such as the Wilson loop $W(t,r)$, are not directly accessible. This challenge has spurred remarkable progress over the last two years, both in terms of concepts and methods, moving us towards a reliable evaluation of Eq.\eqref{Eq:VRealTimeDef} on the lattice. One of the central ingredients, acting as bridge between the Euclidean and Minkowski domain, is the following spectral decomposition \cite{Rothkopf:2009pk,Rothkopf:2011db}
\begin{eqnarray}
 \nonumber W(\tau,r)=\int d\omega e^{-\omega \tau} \rho(\omega,r)\,
\leftrightarrow\, \int d\omega e^{-i\omega t} \rho(\omega,r)= W(t,r).
\end{eqnarray}
Substituted into the definition Eq.\eqref{Eq:VRealTimeDef}, it relates the values of the potential to the real and positive definite spectral function $\rho(\omega,r)$
\begin{align}
\hspace{-0.2cm}V(r)=\lim_{t\to\infty}\int d\omega\, \omega e^{-i\omega t} \rho(\omega,r)/\int d\omega\, e^{-i\omega t} \rho(\omega,r). \label{Eq:PotSpec}
\end{align}

We face two challenges when trying to obtain the values of the potential in practice. One is related to the fact that extracting the continuous function $\rho(\omega,r)$ from a finite and noisy set of lattice datapoints $W(\tau_n,r),~n=1..N_\tau$ is an inherently ill-defined problem. Indeed a simple $\chi^2$ fit would lead to an infinite set of degenerate spectra that all reproduce the datapoints within their error bars. In order to give meaning to such a task, we turn to the concept of Bayesian inference, an approach well established in mathematical statistics. It provides the formal framework to incorporate additional, so called prior information, such as the positivity of $\rho(\omega,r)$ or smoothness properties, in order to regularize an otherwise under-determined $\chi^2$ fit. For our study we deploy a recently developed Bayesian prescription to spectral function reconstruction \cite{Burnier:2013nla}, which has been shown in mock data studies to give better accuracy and precision than both the conventional Maximum Entropy Method \cite{Asakawa:2000tr}, as well as the implementation with extended search space \cite{Rothkopf:2011ef}.

Carrying out the late time limit in Eq.\eqref{Eq:PotSpec} represents the second challenge. Simply inserting the reconstructed spectrum is impractical, as the numerical $t\to\infty$ limit requires a precision of $\rho(\omega,r)$ currently not obtainable. Instead \cite{Rothkopf:2011db} proposed that, since the dynamics of the Wilson loop at late times will eventually be dominated by the lowest lying peak in $\rho(\omega,r)$, its functional form should be fitted from the lattice spectra. The Fourier transform of Eq.\eqref{Eq:PotSpec} can subsequently be carried out analytically without difficulty. The question of which functional form to use in such a fit however remained and was only answered later in \cite{Burnier:2012az}, where the effects of different time-scales contributing to $\rho(\omega,r)$ were disentangled. Based on the symmetries of the Wilson loop it was shown in a general argument that the lowest lying peak will take the form of a skewed Lorentzian:
\begin{align}
\rho(\omega,r) \propto\frac{|{\rm Im} V(r)|{\rm cos}[{\rm Re}{\sigma_\infty}(r)]-({\rm Re}V(r)-\omega){\rm sin}[{\rm Re} {\sigma_\infty}(r)]}{ {\rm Im} V(r)^2+ ({\rm Re} V(r)-\omega)^2}+{c_0}(r)+{c_1}(r)({\rm Re} V(r)-\omega)+\ldots\,. \label{Eq:SkewLor}
\end{align}
The fitted position thus corresponds to the real part of the potential, the fitted width to the imaginary part. Taking the skewness of the Lorentzian peak into account is crucial, as otherwise e.g. the peak position is systematically overestimated.

\begin{figure}[t]
\centering \vspace{-0.6cm}
 \includegraphics[scale=0.6]{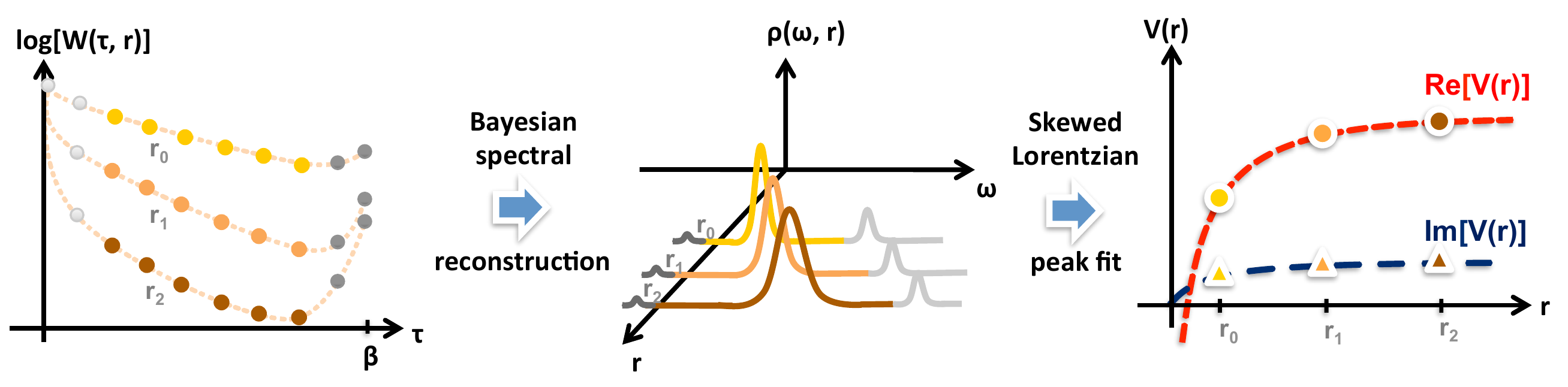} \vspace{-0.3cm}
 \caption{From Euclidean lattice QCD correlators to the complex valued static heavy quark potential (figure adapted from \cite{Rothkopf:2013ria}).}\label{Fig1} \vspace{-0.3cm}
\end{figure}

The viability of this strategy, sketched in Fig.\ref{Fig1}, to access the static potential from Euclidean correlator data has been successfully tested in \cite{Burnier:2013fca} using HTL resummed perturbation theory. At the same time indications were found that for the sole purpose of determining the potential, one can replace the Wilson loop with the correlator of Wilson lines $W_{||}(\tau,r)$ in Coulomb gauge. The latter is devoid of the cusp divergences of the Wilson loop (see \cite{Berwein:2012mw} and references therein) and hence is much easier to measure on the lattice. 

These conceptual and technical improvements combined, allow us to reliably extract the temperature dependence of the static inter-quark potential from quenched and dynamical lattice QCD. Besides presenting our main results \cite{Burnier:2014ssa} we show explicitly the default model dependence of the reconstructions. (For plots of the effects of a removal of datapoints along $\tau$ see \cite{Burnier:2014tbd}).

\section{Quenched Lattice QCD}

\begin{table}[b!]
\begin{tabularx}{15cm}{ | c | X | X | X | X | X | X | X | X | X | }
\hline
	SU(3):$N_\tau$ & 24 & 32 & 40 & 48 & 56 & 64 & 72 & 80 & 96 \\ \hline
	$T$[MeV] & 839 & 629 & 503 & 419 & 360 & 315 & 280 & 252 & 210 \\ \hline
	$N_{\rm meas}$ & 3270 & 2030 & 1940 & 1110 & 1410 &1520 & 860 & 1190 & 1800 \\ \hline
\end{tabularx}
\caption{Quenched SU(3) on $32^3\times N_\tau$ anisotropic $\xi_b=3.5$ lattices with $a_s=0.039$fm and  $T_c\approx271$MeV.}\label{Tab:SU3LatParm}
\end{table}

The first part of our study investigates the real and imaginary part of the static potential in a purely gluonic medium. The reason is practicability, since the relatively low cost of quenched lattice QCD simulations allows us to use lattices with a temporal extend larger than $N_\tau=24$. This is a prerequisite for the reliable determination of spectral peak positions (Re[V]), as well as for a robust order of magnitude estimate of the spectral width, i.e. Im[V], in the presence of our fine lattice spacing. 

Based on the naive anisotropic Wilson action with bare parameters $\beta=7$ and $\xi=3.5$ \cite{Asakawa:2003re} we performed Monte Carlo simulations on $32^3\times N_\tau$ lattices. In this fixed scale approach temperature is varied between $839{\rm MeV} (3.11T_c) \geq T\geq 210 {\rm MeV} (0.78T_c)$ by changing the temporal extend between $24 \geq N_\tau \geq 96$, as tabulated in Tab.\ref{Tab:SU3LatParm}. The physical lattice spacing of $a_s=0.039$fm allows us to resolve the Coulombic region below $r<0.15$fm and gives access to distances where the potential is already screened at $T=839$MeV. Each configuration is fixed to Coulomb gauge iteratively before the Wilson line correlators are measured along each spatial axis, on the square- as well as on the cubic diagonals. We correct for some of the lattice spacing artifacts in the spatial distances by comparing the behavior of free propagators on the lattice to that in the continuum, as described in \cite{Necco:2001xg}.

With the Euclidean Wilson lines at hand, we carry out the spectral reconstruction using all datapoints except those at $\tau=0, \beta$, to avoid possible overlap divergences \cite{Berwein:2012mw}. We discretize the frequency interval between $\omega^{\rm num}\in[-168,185]\times N_\tau/24$ GeV by $N_\omega=4000$ points. $N_{\rm hr}=550$ of those are used to resolve the lowest lying peak in more detail. A flat default model $m(\omega)={\rm const}$ is selected, not to introduce additional bias in the functional form of the end result. Our numerical implementation uses the LBFGS minimizer with high precision arithmetic ($512$ bits) for the task of finding the unique Bayesian spectrum related to the measured data. The solution is accepted if a relative step size of $\Delta=10^{-60}$ is reached. 

\begin{figure}[t]
\centering 
 \includegraphics[scale=0.44]{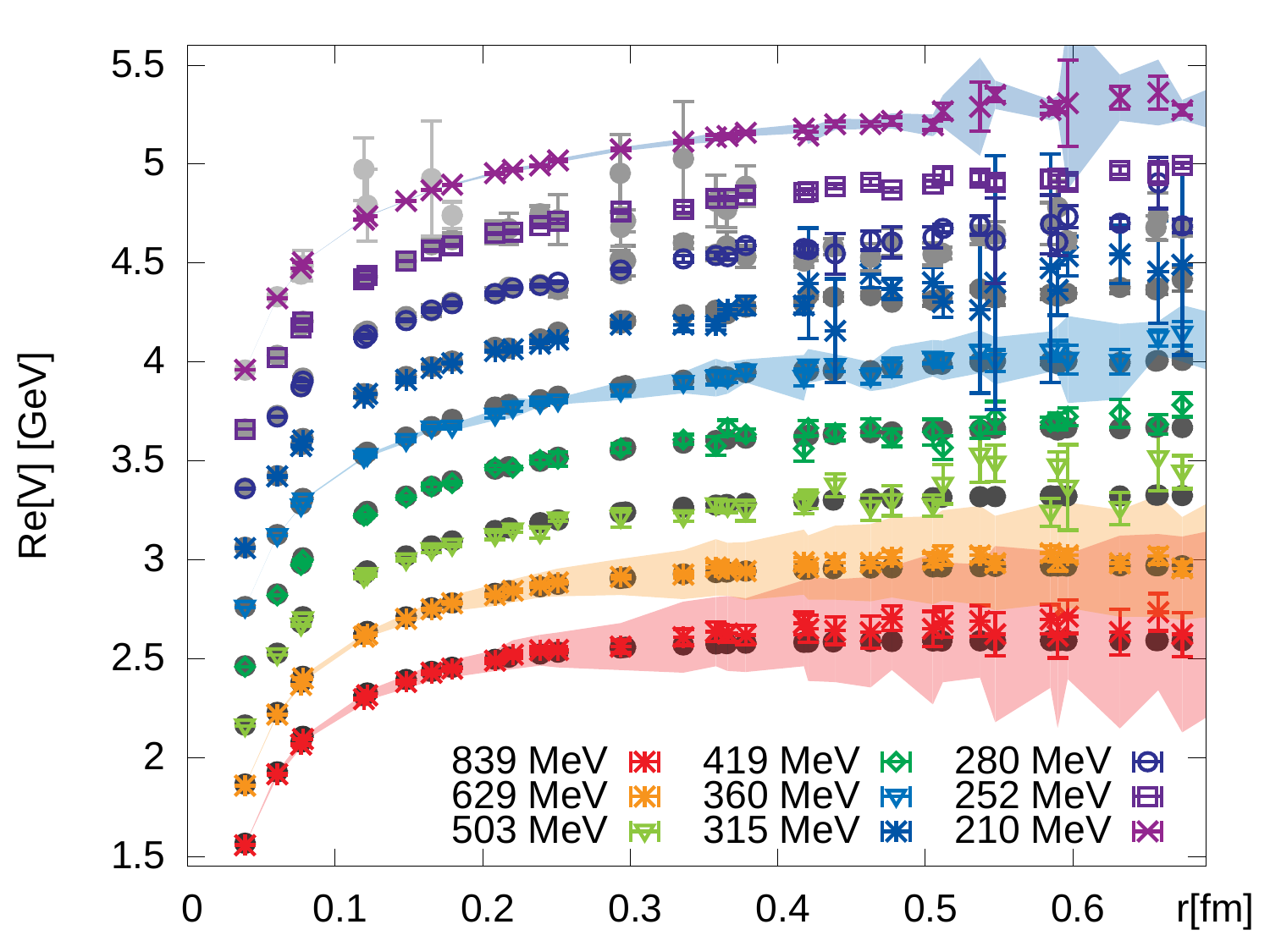} \includegraphics[scale=0.44]{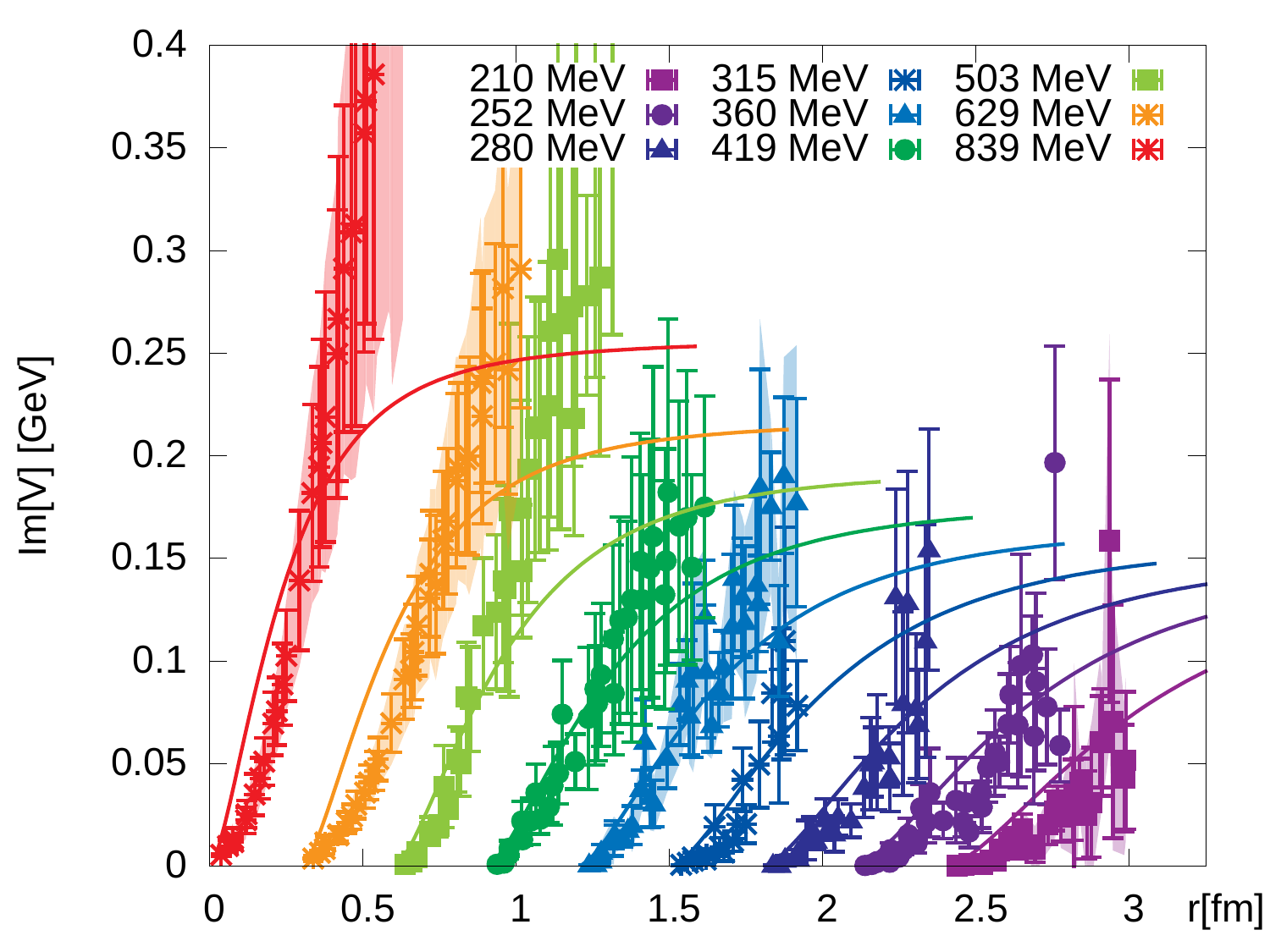}
 \caption{Quenched QCD: (left) Real part of the static inter-quark potential (open symbols) and the color singlet free energies (gray circles), shifted vertically for better readability. The attached error bars derive from the Jackknife variance, error bands are obtained from additional systematics as described in the text. (right) ${\rm Im}[V]$ (symbols) compared to predictions from leading order hard-thermal loop perturbation theory (solid lines).}\label{Fig2} 
\end{figure}

The values of the real and imaginary part shown in Fig.\ref{Fig2} are obtained from a subsequent fit of the full width at half maximum of the reconstructed lowest lying peak with the skewed Lorentzian form of Eq.\ref{Eq:SkewLor}. $\sigma_\infty(r)$ and the $c_i(r)$'s are treated as fit parameters. The error bars are obtained from the Jackknife variance between ten reconstructions, each excluding  a consecutive block of 10\% of the measurements $N_{\rm meas}$. 

\begin{figure}[t]
\centering 
 \includegraphics[scale=0.44]{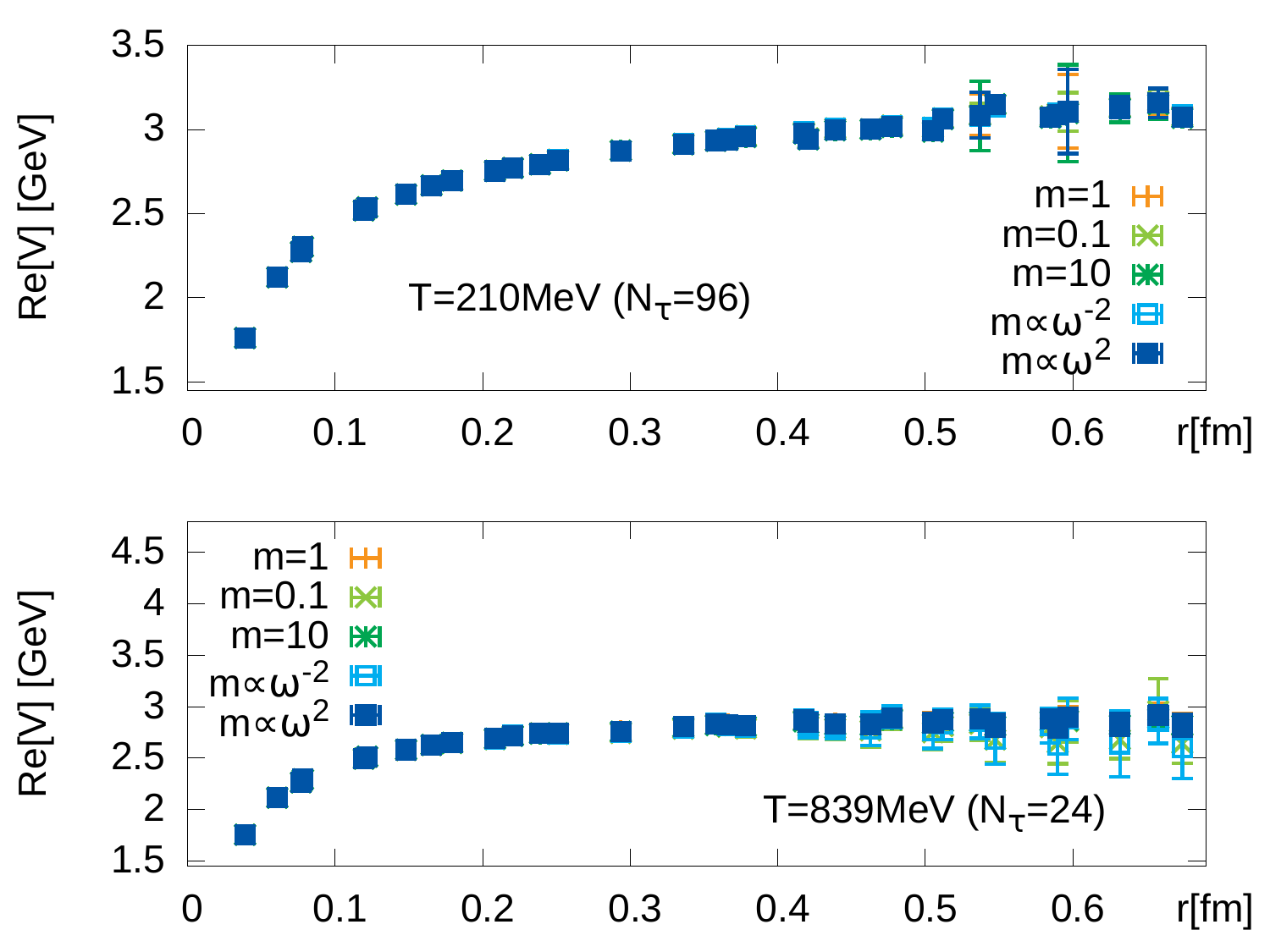} \includegraphics[scale=0.44]{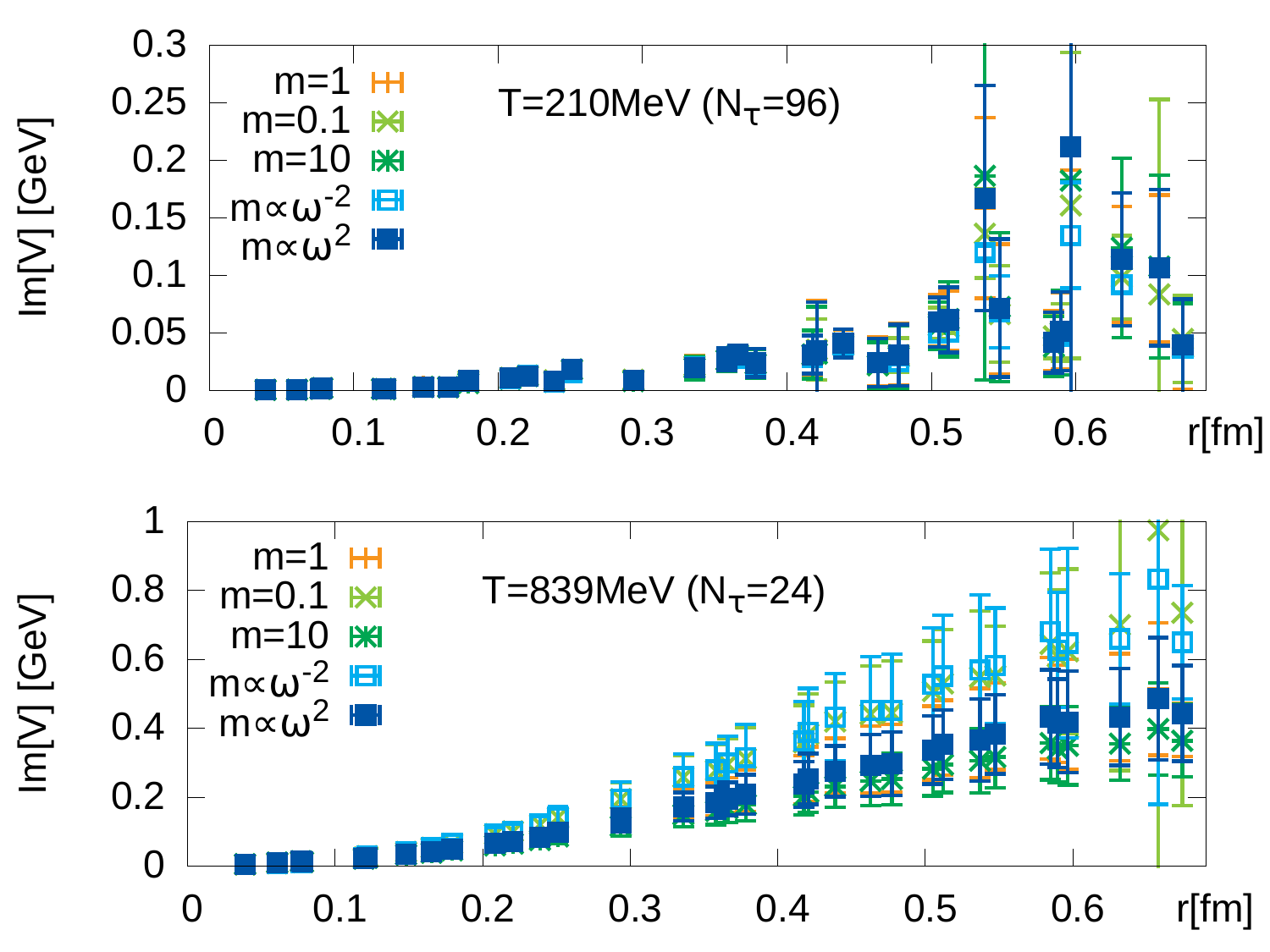}
 \caption{Quenched QCD: Differences in Re[V] (left) and Im[V] (right) from varying the default model normalization ($\times 10$, $\times 0.1$) and functional form ($m\propto{\rm const},\omega^{-2},\omega^2$). At $N_\tau=96$ due to the abundance of datapoints and relatively high statistics, neither the real nor the imaginary part are susceptible to changes of $m(\omega)$. On the other hand for $N_\tau=24$ at $r>0.5$fm the real part shows a weak dependence, while the imaginary part varies visibly from $r>0.35$fm on. }\label{Fig3} 
\end{figure}
We find that the real part of the potential weakens gradually. I.e. the confining linear rise present at low temperatures transitions smoothly to a Debye screened behavior in the deconfined phase. Even though we excluded the $\tau=\beta$ datapoint in the reconstruction of the potential, Fig.\ref{Fig2} shows that the values of Re[V] (open colored symbols) lie close to the color singlet free energies in Coulomb gauge $F^{(1)}(r)=-T{\rm log}[W_{||}(r,\tau=\beta)]$ (gray filled circles). Note that perturbation theory indicates \cite{Burnier:2009bk} that at intermediate distances Re[V] will lie slightly below $F^{(1)}(r)$. The spectral determination of Re[V] at temperatures around the phase transition shows much less variation compared to the free energies, since it does not only rely on information at $\tau=\beta$ but takes into account multiple datapoints along $\tau$. From the fit of the spectral width we find that Im[V] takes on values that are of the same order of magnitude as those predicted by HTL resummed perturbation theory. This first principles result for the imaginary part is consistent with a recent study \cite{Bazavov:2014kva} which fitted the Wilson line correlator using HTL spectral functions.

Simply quoting the statistical errors on the potential is not sufficient to capture the uncertainties, which enter from the underlying spectral reconstruction. Hence our study attempts to quantify also the systematic effects, which are given as error bands in Fig.\ref{Fig2}. They are estimated using the maximum variation arising from three different changes. First is a reduction of the number of datapoints along $\tau$ by four or eight, second the change of the default model normalization ($\times 10$, $\times 0.1$) or functional form ($m\propto{\rm const},\omega^{-2},\omega^2$) and last the removal of 10\%, 20\% or 30\% of the statistics of the input data. Let us discuss the first two factors in more detail.

In Fig.\ref{Fig3} we show the dependence of the potential reconstruction on changes to the default model at the lowest ($T=210$MeV, $N_\tau=96$) and highest ($T=839$MeV, $N_\tau=24$) temperature. At $N_\tau=96$ we find that, due to the abundance of available datapoints and the relatively high statistics, both Re[V] and Im[V] are very robust and do not show variation beyond the small statistical error bars. In the case of much fewer datapoints ($N_\tau=24$), the real part still remains stable, while the spectral width shows a dependence to the form of $m(\omega)$ for $r>0.3$fm. Note however that this reduction in reconstruction reliability is accompanied with an increase in the Jackknife variance, so that these error bars overlap for all tested scenarios.
 
\begin{figure}[t]
\centering
 \includegraphics[scale=0.44]{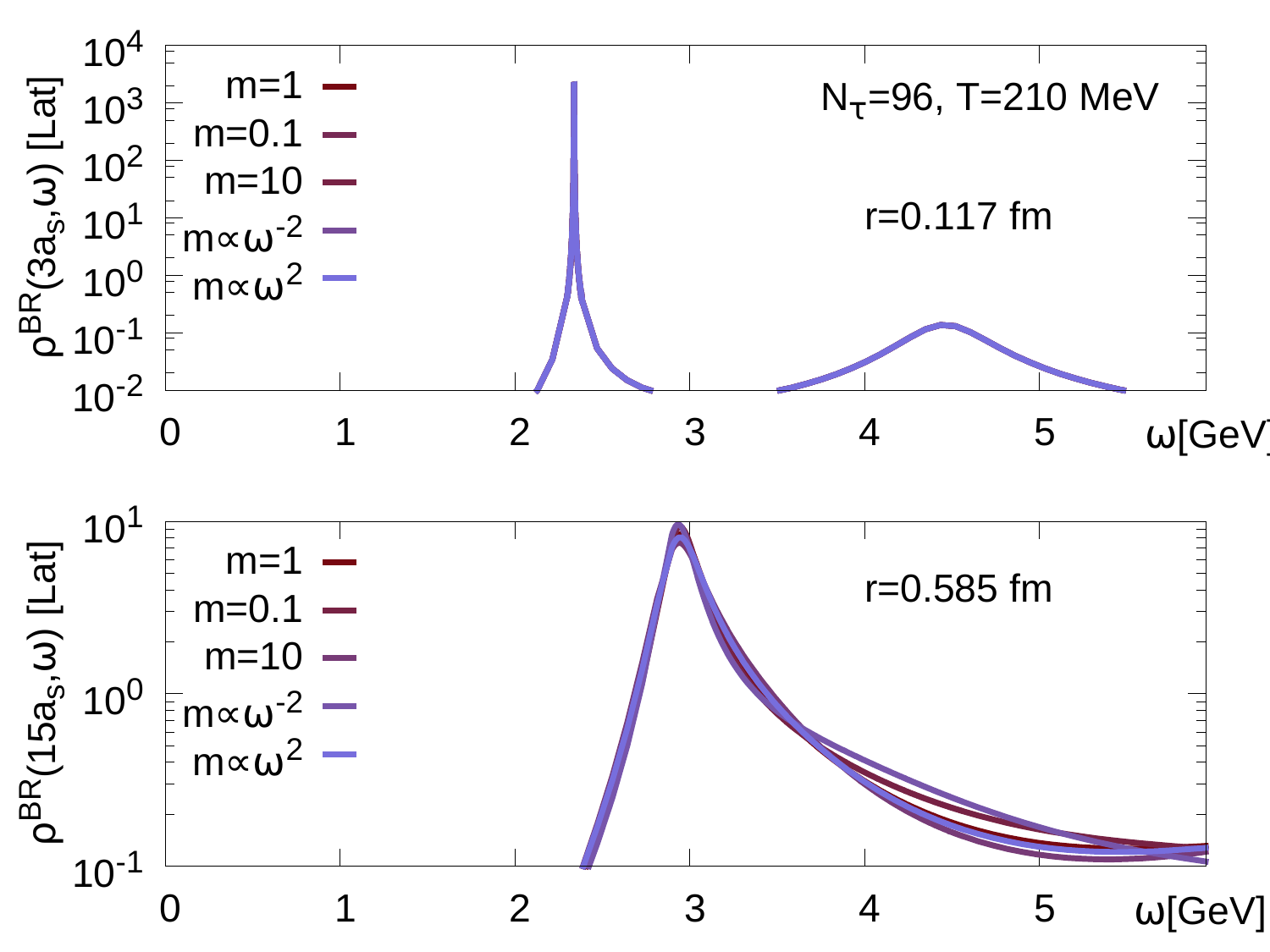}\includegraphics[scale=0.44]{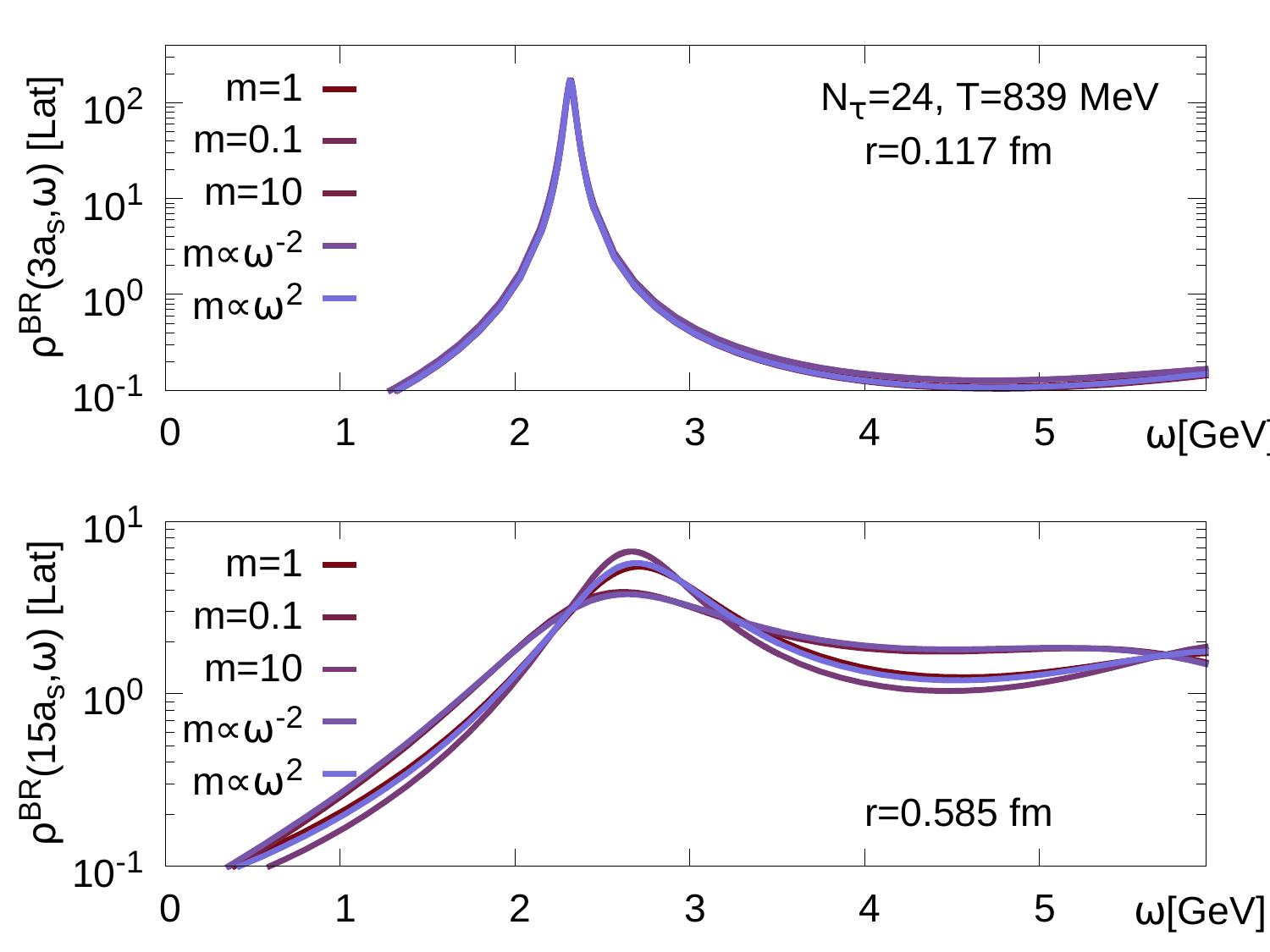}
 \caption{Quenched QCD: Default model dependence of the reconstruction of the Wilson line spectra at (left) $T=360$MeV and (right) $T=839$MeV for two different spatial separation distances (top) $r=0.117$fm and (bottom) $r=0.585$fm. At $r=3a_s$ the position and width of the peak are insensitive to the changes in $m(\omega)$ at both temperatures. For the larger separation ($r=15a_s$) the higher temperature reconstruction shows more variation due to the small number of available data points.}\label{Fig4}
\end{figure}

The explicit form of the reconstructed spectra at small and intermediate spatial separation distances and their variation with $m(\omega)$ is shown in Fig.\ref{Fig4}. As the form of the lowest lying peak structure constitutes the basis for the potential extraction, the behavior of Re[V] and Im[V] observed in Fig.\ref{Fig3} is reflected in these plots. The influence of $m(\omega)$ on the position and width of the peak at $r=0.117$fm is insignificant at both temperatures. The larger number of available datapoints for $N_\tau=96$ keeps the reconstruction of the lowest peak stable even at $r=0.58$5fm, while for $N_\tau=24$, variation in the peak position and width can already be identified by eye.

We have also checked what happens to the result of the potential extraction if the number of datapoints is diminished. While this systematic effect dominates the error bands at high temperatures, it is reassuring that at $N_\tau=96$, taking into account only the first $\tau_{\rm max}=32$ points still leads to the same result for Re[V] within statistical errors. I.e. simply removing the late $\tau$ data does not mimic the effects an actual increase in temperature (to $N_\tau=32$, $T=629$MeV) would have. In the case of $N_\tau=24$ on the other hand, reducing the number of datapoints, produces significant changes in the values we obtain for Re[V]. The decrease in reliability of the reconstruction however is captured well in the simultaneous increase of the Jackknife variance, so that the best result still lies within the error bars. These observations give us confidence that the changes in Re[V] can indeed be attributed to the effects of a thermal medium and not simply to a degradation of the spectral reconstruction at higher T.

Im[V] reacts to the removal of datapoints even at $N_\tau=96$. Its values appears to decrease, as can be expected. The width of the lowest lying peak dominates the correlator exactly at the late $\tau$ values we discard in this example. At intermediate imaginary time on the other hand its contribution is exponentially suppressed and is not easily extracted. Note that we are still able to obtain a finite results for Im[V] at $N_\tau=24$ if the correlator values close to $\tau=\beta$ are retained.

\section{Dynamical Lattice QCD}

\begin{table}[b!]
\begin{tabularx}{15cm}{ | c | X | X | X | X | X | X | X| }
\hline
	QGP: $\beta$ \hspace{0.8cm}& 6.8 & 6.9 & 7 & 7.125 & 7.25 & 7.3 & 7.48 \\ \hline
	$T$[MeV] & 148 & 164 & 182 & 205 & 232 & 243 & 286 \\ \hline
	a [fm] & 0.111 & 0.1 & 0.09 & 0.08 & 0.071 & 0.068 & 0.057\\ \hline
	$N_{\rm meas}$ & 1295 & 1340 & 1015 & 840 & 1220 & 1150 & 1130 \\ \hline
\end{tabularx}
\caption{The isotropic HotQCD $48^3\times12$ lattices with ASQTAD action ($m_l=m_s/20,T_c\approx174$MeV).}\label{Tab:DynLatParm}
\end{table}

The second part of our study is aimed at the extraction of the in-medium potential in a true quark-gluon plasma with light u, d and s quarks. Taking into account the effect of dynamical quarks is essential for phenomenological applications, as it e.g. shifts the (pseudo-) critical temperature by more than $100$MeV. To this end we repeat the potential extraction on lattices generated by the HotQCD collaboration \cite{Bazavov:2011nk} for the study of the equation of state. Their isotropic $48^3\times12$ configurations are based on the $N_f=2+1$ ASQTAD action ($m_l=m_s/20$) and span the temperature range $286 {\rm MeV} (1.64T_c) \geq T\geq 148{\rm MeV} (0.85T_c)$. A fixed box approach is used, so that by varying the lattice spacing $\beta=6.8\ldots7.48$, at each T we have at our disposal the same number of points in temporal direction (see Tab.\ref{Tab:DynLatParm}). The significantly higher computational cost for dynamical quarks currently limits us to the use of $N_\tau=12$, which unfortunately does not allow a reliable determination of spectral widths even with the high precision of the available data.

\begin{figure}[t]
\centering
 \includegraphics[scale=0.44]{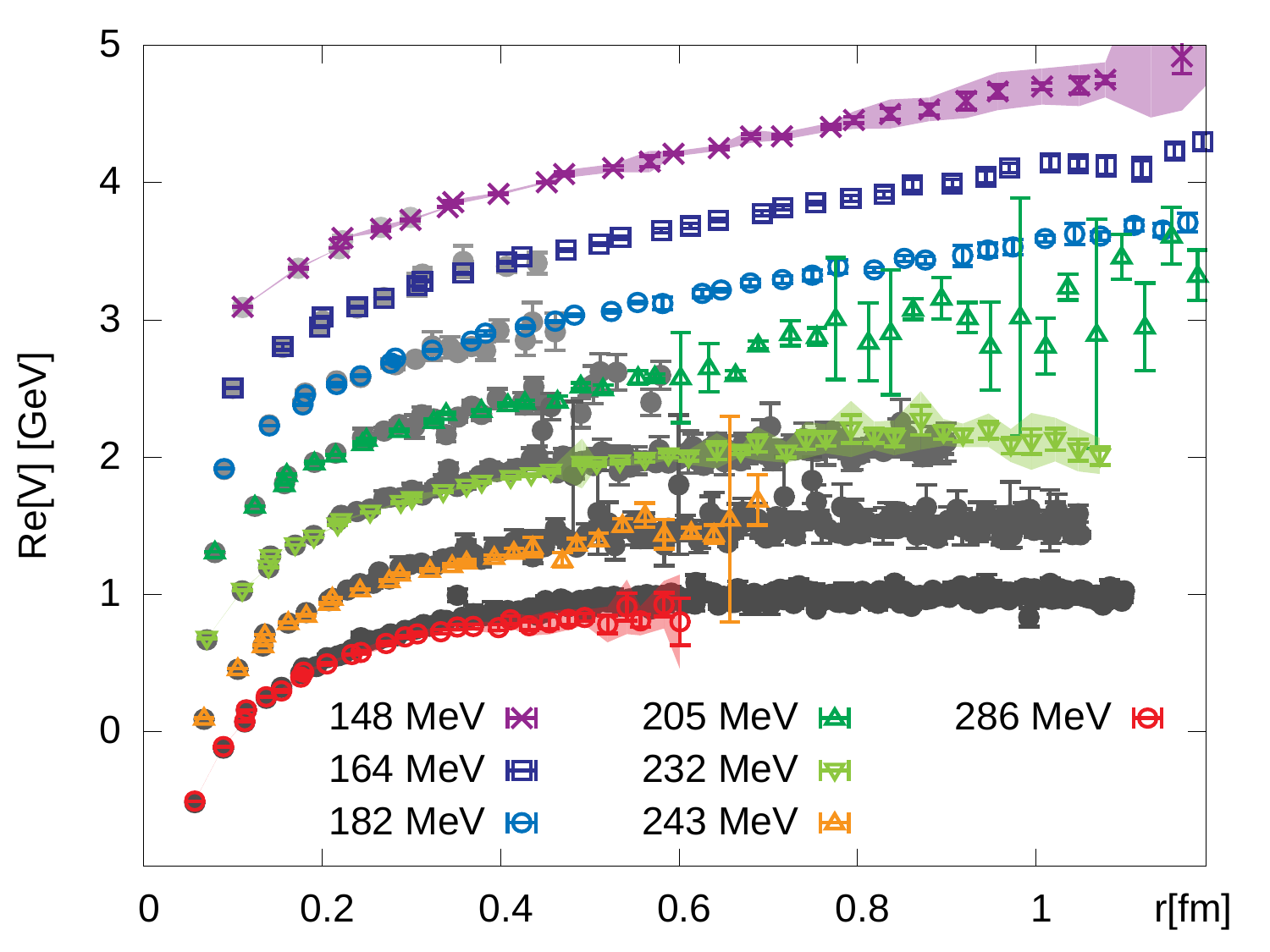}\includegraphics[scale=0.44]{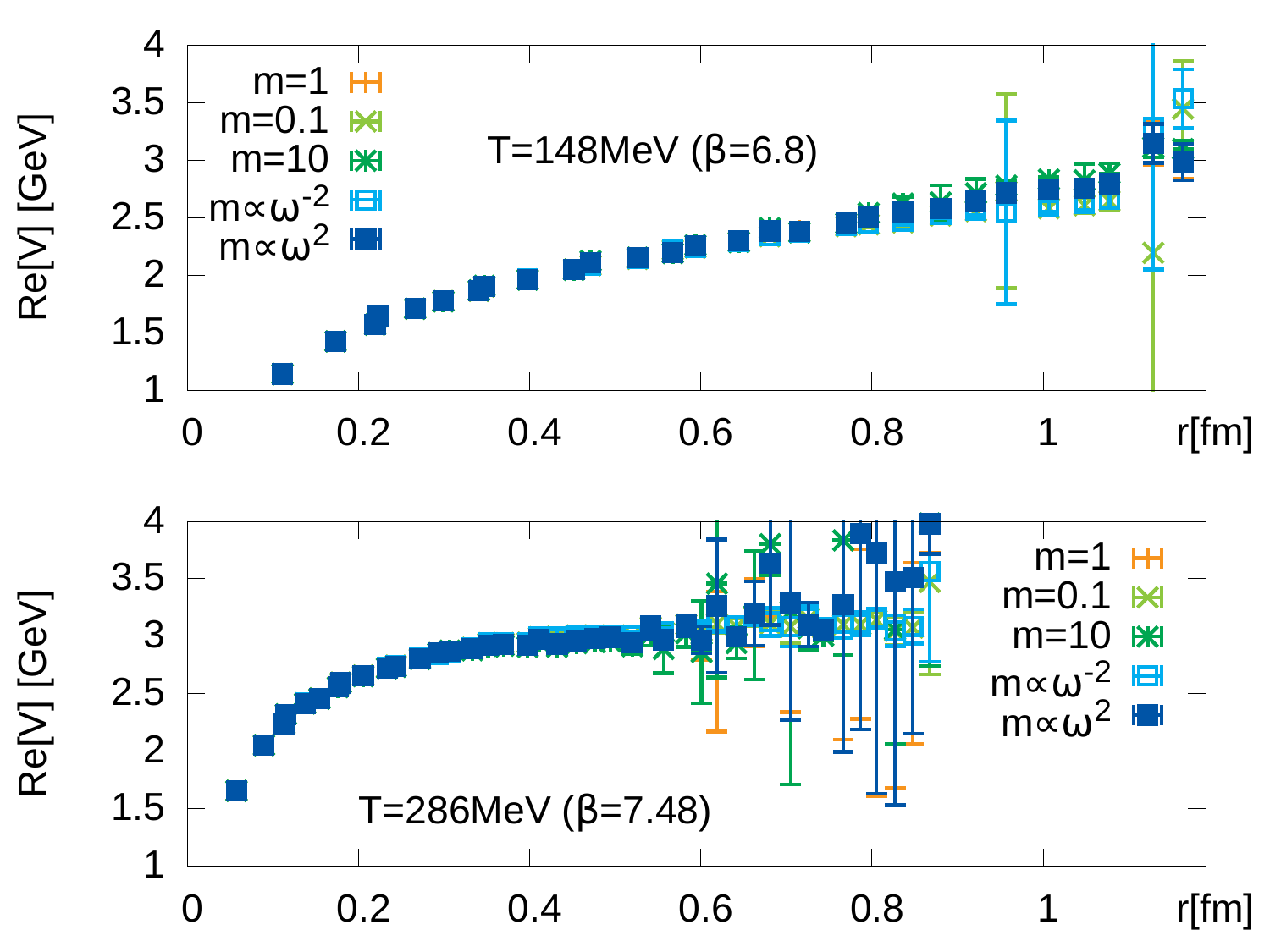}
 \caption{Full QCD: (left) Comparison of the real part of the static inter-quark potential (open symbols) and the color singlet free energies (gray circles). The values are shifted for better readability. (right) Results for Re[V] for different choices of default model normalization ($\times 10$, $\times 0.1$) and functional form ($m\propto{\rm const},\omega^{-2},\omega^2$) at the lowest ($\beta=6.8$) and highest ($\beta=7.48$) temperatures. }\label{Fig5} 
\end{figure}

Since the number of datapoints does not change between temperatures, we carry out the spectral reconstruction with a common choice for $\beta^{\rm num}=20$ and discretize the frequencies between $\omega\in[-11,12]$ in these units. From the overall $N_\omega=4600$ points we use $N_{\rm hr}=1000$ to resolve the lowest lying peak. Fitting this structure at the full-width at half maximum with the functional form of Eq.\eqref{Eq:SkewLor} allows us to extract Re[V] shown in Fig.\ref{Fig5}. 

As already seen in the quenched case, the determination of Re[V] based on a spectral reconstruction is very robust at $T\lesssim T_c$. This allows us to reach distances up to $r\simeq 1.2$fm at which the signal for the free energies is already lost in statistical noise. The real part does not show any signs of string breaking at the separations we investigated, probably due to the still relatively large pion masses of $M_\pi\approx 300$MeV on our lattices. Nevertheless in the presence of dynamical quarks, the Debye screened behavior in Re[V] is already well pronounced at 286MeV. Interestingly the values of the real part again lie close to those of the color singlet free energies in Coulomb gauge.

\begin{figure}[t]
\centering
 \includegraphics[scale=0.44]{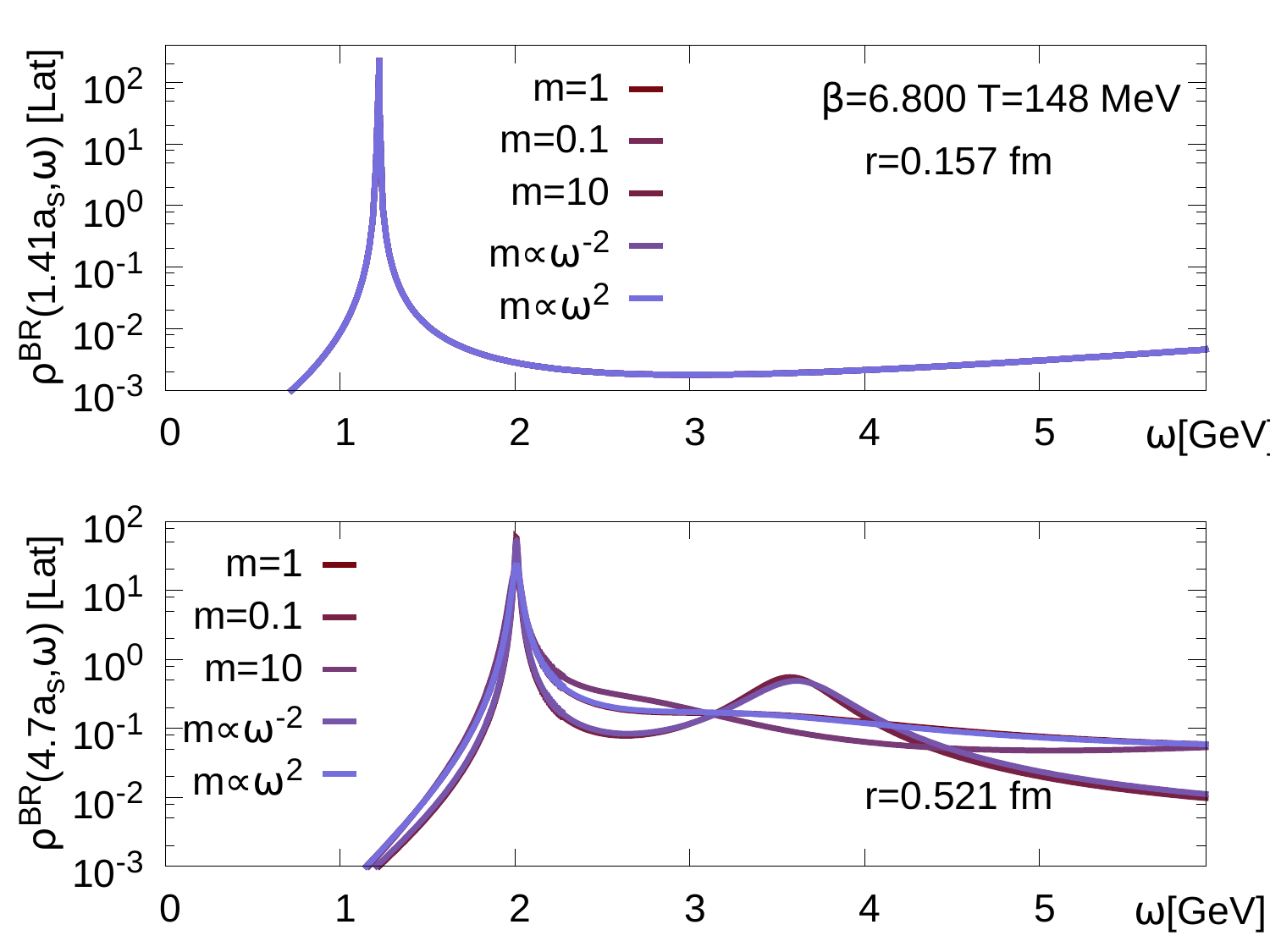}\includegraphics[scale=0.44]{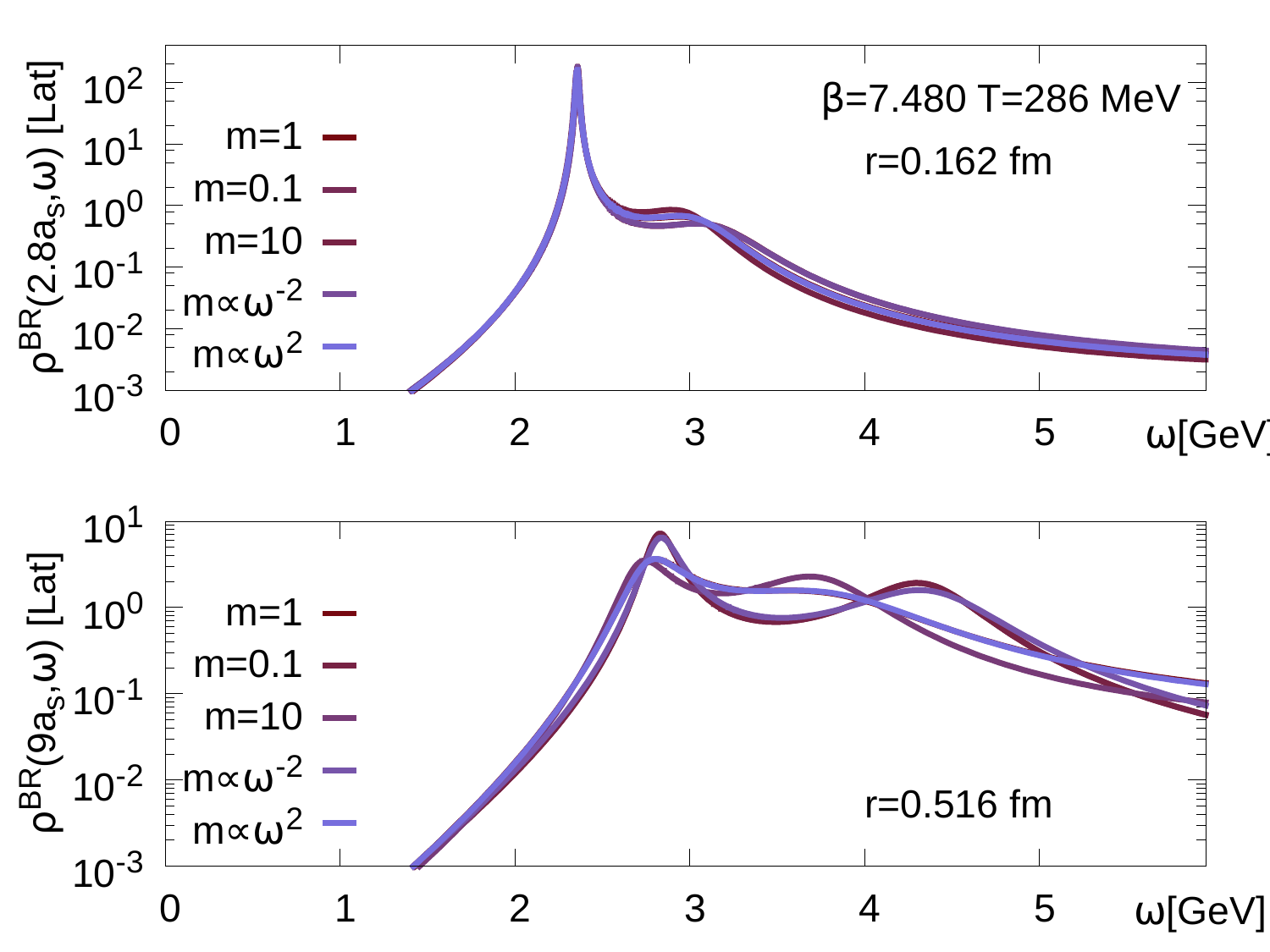}
 \caption{Full QCD: Dependence of the Wilson line spectral reconstruction on changes in the default model at the lowest ($\beta=6.8$, left) and highest ($\beta=7.48$, right) temperatures. At small separation distances $r<0.17$fm the peak position is not affected by the choice of $m(\omega)$. Since $r\simeq0.52$ amounts to $r/a_s=4.7$ at low but already to $r/a_s=9$ at high T, the lowest lying peak in the latter case shows more variation with the default model.}\label{Fig6}
\end{figure}

Statistical uncertainty is represented by the error bars in Fig.\ref{Fig5}, which are obtained from the Jackknife variance between ten different reconstructions, from which consecutive blocks of 10\% of the correlator measurements had been removed. The error bands contain also systematics and represent the maximum variance from either changing the number of available correlator datapoints by one or two, varying the normalization ($\times 10$, $\times 0.1$) or the functional form ($m\propto{\rm const},\omega^{-2},\omega^2$) of the default model or decreasing the statistics of the input data by 10\%, 20\% or 30\%.

One of the systematic uncertainties we need to understand is the dependence of the result on the default model. The two panels on the right of Fig.\ref{Fig5} show explicitly how the extracted values of Re[V] change with $m(\omega)$. While at low temperature ($T=148$MeV) the real part is virtually unaffected at all distances probed, we find that at $T=249$MeV the reconstruction becomes more and more unreliable beyond $r>0.6$fm. The Jackknife variance however captures this degradation of the extraction, in that the size of the error bars grows accordingly.

The influence of different choices of $m(\omega)$ on the spectral reconstruction underlying the potential extraction is shown Fig.\ref{Fig6} at the lowest ($T=148$MeV, $\beta=6.8$) and highest ($T=286$MeV, $\beta=7.48$) temperature. Even with the little number of datapoints available, the reconstructions at the small values of $r<0.17$fm show a stable position of the lowest lying peak, with some variation at higher frequencies visible. While a separation distance of $r\approx0.52$fm corresponds to only $4.7a_s$ at $\beta=6.8$ it already amounts to $9a_s$ at $\beta=7.48$. At the same time a smaller lattice spacing leads to a higher UV cutoff, so that the reconstruction has to capture more structure overall. We thus expect and find in Fig.\ref{Fig6} that for the larger separation distances, the lowest lying peak at $\beta=7.48$ is less reliably reconstructed than at $\beta=6.8$.

In the case of dynamical lattices with a small number of datapoints to begin with, it is also important to quantify the effects of removing further points along $\tau$. Note that discarding a single temporal step at $\beta=6.8$ and $\beta=7.48$ amounts to a different change in the actual time extend due to the difference in physical lattice spacing. With the relatively high statistics available, we find that the extraction of Re[V] is stable against small changes in $\tau_{\rm max}$. This again gives us confidence to attribute the change, from the confining behavior at low T to a Debye screened form at $T>T_c$, to the effects of the thermal medium and not to a degradation of the spectral reconstruction.

\section{Conclusion}

With the help of recent conceptual and technical progress, a reliable determination of the static inter-quark potential from first principles thermal lattice QCD has become possible. A major role is played by a novel Bayesian approach to spectral function reconstruction, which for the first time allows us to faithfully reconstruct the sharp peaked features of the Wilson line spectra from Euclidean time Monte Carlo data. Fitting the position and width of these spectra, we determined the real and imaginary part in quenched QCD and Re[V] on realistic dynamical lattices with light u,d and s quarks. We find a gradual transition of the linear confining behavior to a Debye screened form in Re[V] and observe that its values lie close to those of the color singlet free energies in Coulomb gauge. Our estimate for the imaginary part on quenched lattices takes on the same order of magnitude as the predictions from HTL resummed perturbation theory. We furthermore checked the systematic dependencies of the Bayesian reconstruction including various default models. The spectra at $T\simeq T_c$ are very stable, while the decrease in physical extend or in the number of points in Euclidean time at higher temperatures degrades the outcome at separation distances $r>0.3$fm, as expected.  

\vspace{-0.15cm}
\begin{theacknowledgments}
 The authors thank H.~B. Meyer, M.~P. Lombardo, P. Petreczky and J.-I. Skullerud for fruitful discussions. YB is supported by SNF grant PZ00P2-142524.
\end{theacknowledgments}

\vspace{-0.35cm}
\bibliographystyle{aipproc}   


\end{document}